\documentclass[structabstract]{aa}  
\usepackage{graphicx}
\usepackage{natbib}
\bibpunct{(}{)}{;}{a}{}{,} 
\usepackage{txfonts}
\begin{document}
   \title{The impact of chemical differentiation of white dwarfs on thermonuclear supernovae}
\authorrunning{E. Bravo et al.}
\titlerunning{The impact of chemical differentiation on thermonuclear supernovae}


   \author{E. Bravo\inst{\ref{inst1}}
	  \and
          L. G. Althaus\inst{\ref{inst2},\ref{inst3},\ref{inst4}}
          \and
          E. Garc\'\i a-Berro\inst{\ref{inst2},\ref{inst5}}
          \and
          I. Dom\'\i nguez\inst{\ref{inst6}}
          }

   \institute{Departament de F\'\i sica i Enginyeria Nuclear, 
              Universitat Polit\`ecnica de Catalunya, 
              c/Comte d'Urgell 187, 
              08036 Barcelona, Spain\\   
              \email{eduardo.bravo@upc.edu} \label{inst1}
         \and
              Departament de F\'\i sica Aplicada, 
              Universitat Polit\`ecnica de Catalunya, 
              c/Esteve Terrades 5,
	      08860 Castelldefels, Spain, 
	      \email{garcia@fa.upc.edu} \label{inst2}
         \and
	      Facultad de Ciencias Astron\'omicas y Geof\'\i sicas, 
              Universidad Nacional de La Plata, 
              Paseo del Bosque s/n, 
              1900 La Plata, Argentina
	      \email{althaus@fcaglp.fcaglp.unlp.edu.ar} \label{inst3}
	 \and
	      Instituto de Astrof\'\i sica de La Plata (CCT La Plata), 
              CONICET, 
              1900 La Plata, Argentina \label{inst4} 
	 \and 
	      Institut d’Estudis Espacials de Catalunya, 
              Ed. Nexus-201, c/Gran Capita 2-4, 
              08034 Barcelona, Spain\label{inst5}
         \and
              Departamento de F\'\i sica Te\'orica y del Cosmos, 
              Universidad de Granada, 
              18071 Granada, Spain\label{inst6}
             }

   \date{Received \today; accepted }
 
  \abstract
   {}
   {Gravitational settling  of $^{22}$Ne  in cooling white  dwarfs can
     affect the outcome of thermonuclear supernovae.
     We investigate  how the supernova  energetics and nucleosynthesis
     are  in turn influenced by  this  process.  We use realistic
     chemical  profiles derived from state-of-the-art  white  dwarf
     cooling sequences.  The cooling  sequences provide a link between
     the white dwarf  chemical structure and the age  of the supernova
     progenitor system.}
   {The  cooling  sequence of  a  1~M$_{\sun}$  white  dwarf was computed until freezing 
using an up-to-date stellar evolutionary
     code.    We computed   explosions  of   both
     Chandrasekhar  mass  and   sub-Chandrasekhar  mass  white  dwarfs,
     assuming spherical symmetry and neglecting convective mixing during the pre-supernova carbon
simmering phase to maximize the effects of chemical separation.}
   {Neither   gravitational  settling   of   $^{22}$Ne  nor   chemical
     differentiation  of  $^{12}$C and  $^{16}$O  have an  appreciable
     impact on the properties of Type Ia supernovae, unless there is a
     direct dependence of the  flame properties (density of transition
     from deflagration to detonation)  on the chemical composition. At
     a  fixed  transition  density, the maximum variation in the supernova magnitude obtained from
progenitors of different ages is $\sim0.06$ magnitudes, and even assuming an
unrealistically
large diffusion coefficient of $^{22}$Ne it would be less than $\sim0.09$~mag.  However, if  the 
transition density  depends on  the
     chemical composition  (all other  things being equal)  the oldest
     SNIa can be as much  as 0.4 magnitudes brighter than the youngest
     ones (in our models the  age difference is 7.4 Gyr). In addition, our
     results show  that $^{22}$Ne  sedimentation cannot be  invoked to
     account  for   the  formation  of   a  central  core   of  stable
     neutron-rich Fe-group  nuclei in the  ejecta of sub-Chandrasekhar
     models, as  required by observations  of Type Ia  supernovae.} 
   {}


   \maketitle


\section{Introduction}

Thermonuclear explosions  of accreting white dwarfs are  thought to be
the origin of several of  the most violent phenomena known: Type Ia
supernovae   (SNIa)    \citep{hoy60,bran95,hil00,rop06c},   .Ia   supernovae
\citep{bil07},  and even  some Type  Ib supernovae  \citep{per10}. The
discovery   of    the   accelerated   expansion    of   the   Universe
\citep{rie98,per99}  has  put  SNIa  at  the center  of  attention  of
cosmologists,  as a  complete knowledge  of any  systematics affecting
SNIa  luminosity is  necessary to  achieve the  precision  required to
measure   the   equation  of   state   of   dark  matter   \citep[see,
e.g.][]{ton05}.

One of the most  intriguing systematics concerning SNIa variability is
the  correlation  suggested  by  the  observations  between  supernova
brightness   and  progenitor   age,  the   so-called   ``age  effect''
\citep[there is  another type of  age effect that correlates  the SNIa
{\sl rate}  with age,  see for instance][]{nom00}.  The  luminosity of
SNIa is  related to  the morphological types  of their  host galaxies:
brighter SNIa  tend to occur  in spiral galaxies with  younger stellar
populations, while most of  the fainter events occur preferentially in
early-type      galaxies     with      relatively      older     stars
\citep{ham96,ham00,wan06}.  This observational finding implies that the age
difference  of  the  progenitors  is   one  of  the  origins  of  SNIa
diversities.  
The age effect may have its origin in either the intrinsic dependence of the observational
properties of individual supernovae on the age of their progenitors (this could be
called a pure age effect) or the occurrence of different channels leading to SNIa, each one
characterized by slightly different properties and time histories (e.g. white dwarf mergers versus
single degenerate progenitors), or even in observational biases. In this work, we focus on a
possible pure age effect caused by the chemical differentiation that occurs during white dwarf
cooling prior to the supernova event.

Observational constraints on the nature of the age effect come from statistical studies of
the properties of large numbers of SNIa in different environments.
\citet{gal08} found a strong correlation suggesting that
SNIa in galaxies whose population  is older than 5 Gyr are $\sim1$~mag
fainter  at  maximum  than   those  found  in  galaxies  with  younger
populations.  They concluded that  the time since progenitor formation
primarily determines the production of $^{56}$Ni, the main radioactive
nuclide powering the light curve. A similar conclusion was obtained by
\citet{how09}.   \citet{nei09}  pointed  out  what  appears  to  be  a
threshold average population  age of 3~Gyr above which  a host is less
likely   to  produce   SNIa   with  $^{56}$Ni   masses  greater   than
$\sim0.5$~M$_{\sun}$.  Below  this age threshold, there  appears to be
little  correlation between  $^{56}$Ni mass  and host  age. 
These findings tightly constrain the possible origin of the age effect. 
For instance, \citet{kru10} attributed the luminosity-age correlation to an increase
in the central density of the white dwarf prior to the SNIa explosion, during the accretion
phase  of the  progenitor system.  However,  they  only explored
cooling  times  below  $\sim1$~Gyr,  which  are  much  less  than  the
threshold suggested by \citet{nei09}.

\citet{tim03} and \citet{ton05} argued that sedimentation of $^{22}$Ne
over a timescale  of several (up to $\sim7-8$)  Gyr provides a natural
time-dependent mechanism  modulating the  luminosity of a  white dwarf
explosion.   The  sequence  of   events  can  be  divided  into  three
phases. First, the primary component of a binary system evolves into a
white  dwarf. For  the range  of  white dwarf  progenitor masses,  the
timescale    of   the    first   phase    is   on    the    order   of
$\lesssim1$~Gyr.  Second, there is  a variable  period of  time during
which the secondary  remains in the main-sequence and  the white dwarf
cools  and acquires  a (perhaps  partially)  chemically differentiated
structure.  Third, there is an  accretion phase during which the white
dwarf  grows  in  mass  and  increases its  central  density.  At  the
accretion rates necessary  to allow a SNIa explosion,  the duration of
this last phase  can be much less than 1~Gyr.   For nearby SNIa, which
presumably  sample a range  of progenitor  ages, the  sedimentation of
$^{22}$Ne during the second phase might introduce a larger variability
of luminosity than for SNIa exploding at higher redshift, for
instance $z\sim1$.

From  the  theoretical point  of  view, there  is  a  long history  of
calculations  that have tried  to unravel  the influence  of $^{22}$Ne
sedimentation   on   the   physics  of   SNIa   \citep[e.g.][]{bra92}.
\citet{bil01} proposed to  use the production of $^{54}$Fe  in SNIa as
an  indirect test  of the  sedimentation of  $^{22}$Ne.  \citet{pir08}
analyzed  the impact  of $^{22}$Ne  sedimentation on  the size  of the
convective zone during  the last stages of carbon  simmering, which set
the   initial    conditions   for   a    SNIa   explosion   \citep[see
also][]{pir08b}. They  claimed that  to have an  appreciable effect it
would  be necessary  for all  the  $^{22}$Ne to diffuse into  the
convective  core prior  to carbon  ignition. In their modeling of
sub-Chandrasekhar white dwarf  explosions, \citet{sim10} had to resort
to  a hypothetical gravitational  settling of  $^{22}$Ne to reproduce the
central  concentration   of  neutron-rich  isotopes inferred from
observations of SNIa.

\citet{gar08},     \citet{alt10},    \citet{ren10},    and
\citet{gar10} computed realistic  white dwarf  cooling sequences
with   updated  physics,   in   which  the   evolution  was   followed
simultaneously  with  the  several possible  chemical  differentiation
processes: gravitational settling of $^{22}$Ne during the liquid phase
and chemical  separation during the liquid-solid  phase transition. It
is, thus,  interesting to  know the consequences of these
state-of-the-art white dwarf evolutionary calculations on the
outcome of  the thermonuclear  explosions of these objects, to  see whether
they meet  the expectations  raised by the  above-mentioned  models of
SNIa.  The  plan of the  paper is as  follows. In the next  section, we
briefly outline the chemical differentiation processes that take place
during  white dwarf  cooling. The  chemical profiles  obtained  at both the
beginning and  the  end of  the cooling process  are used  as input
models for the  thermonuclear supernova code. In Section~\ref{sectddt},
we report the  results of the explosion of  a Chandrasekhar-mass white
dwarf, and the  sensitivity of the supernova properties  to the age of
the  progenitor.   Section~\ref{sectsub} is  devoted  to evaluating the
impact  of  the chemical  profiles  on  sub-Chandrasekhar white  dwarf
explosions,  where  we  account  for  their  possible  impact  on  the
formation of a carbon detonation  at the top of the carbon-oxygen core
as  a  consequence  of  a  helium  detonation  (the  double-detonation
scenario   for   sub-Chandrasekhar   white   dwarfs).    Finally,   in
Section~\ref{sectcon} we summarize our  major findings and draw our
conclusions.

\section{Chemical differentiation during white dwarf cooling}
\label{sectchem}

\begin{figure}[tb]
\centering
   \includegraphics[width=9 cm]{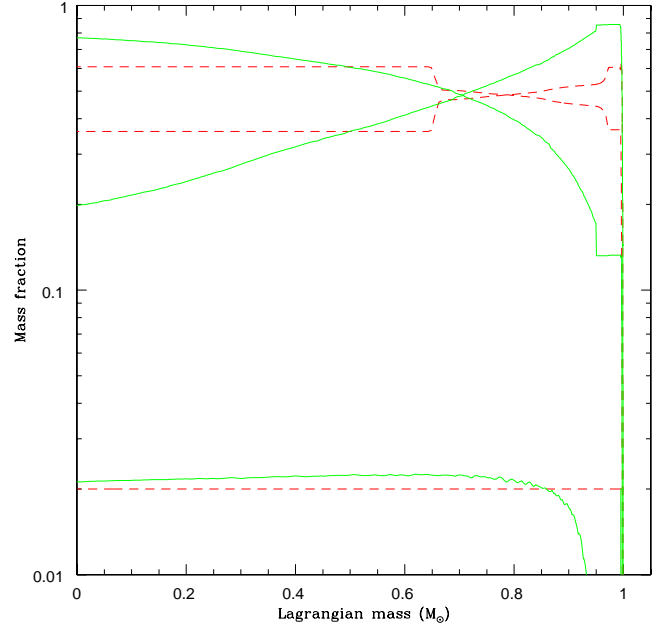}
\caption{Chemical profile of  the white dwarf at the  beginning of the
  cooling   process  (red)   and  the   the  end   of  crystallization
  (green). The three abundances shown  for each model belong (from top
  to bottom at $M=0$) to:  $^{16}$O, $^{12}$C, and $^{22}$Ne. The mass
  fraction of $^{12}$C  in the outermost layer is  0.61 (red line) and
  0.86 (green line).}
\label{fig1}
\end{figure}

The  white dwarf  evolutionary  models  used in  this  work were
extracted from the full evolution of a 1.0~M$_{\sun}$ cooling sequence
computed   with   the   {\tt   LPCODE}   stellar   evolutionary   code
\citep[see][for   details]{alt10}.   Particular  attention was
devoted to the  treatment of the abundance changes  resulting from the
various  physical processes  acting during  the cooling  phase, namely
element diffusion in the  outer layers, carbon-oxygen phase separation
upon core crystallization, and the slow $^{22}$Ne sedimentation in the
liquid regions.   The last process is particularly relevant to the present
work.  $^{22}$Ne  sedimentation towards the center of  the white dwarf
results from an imbalance  between gravitational and electrical forces
caused by their being two  extra  neutrons in the  $^{22}$Ne  nucleus
relative  to  $A_i=2Z_i$  nuclei  \citep{bra92,del02,gar08}.   In  our
simulations, time-dependent  $^{22}$Ne sedimentation was treated
in a self-consistent way with  the white dwarf evolution, and we refer
the reader to \citet{alt10} for details. In particular, for the liquid
regions, we adopt the diffusion coefficient $D= 7.3 \times 10^{-7} T /
\varrho^{1/2}   \overline{Z}   \   \Gamma   ^{1/3}$~cm$^2   ~$s$^{-1}$
\citep[see][]{del02}.   The chemical  stratification  of our  starting
white  dwarf  configuration at  the  beginning  of  the cooling  track
(depicted in Fig. \ref{fig1}) is  the result of the complete evolution
of an  initially 5.0~M$_{\sun}$ model  star evolved from  the zero-age
main  sequence, through  the core  hydrogen-burning phase,  the helium
burning  phase,  and the  thermally  pulsing  asymptotic giant  branch
phase, to the white dwarf stage \citep[see][]{ren10}. The initial mass
abundance of $^{22}$Ne in the core is 0.02 (by mass).

\begin{figure}[tb]
\centering
   \includegraphics[width=9 cm]{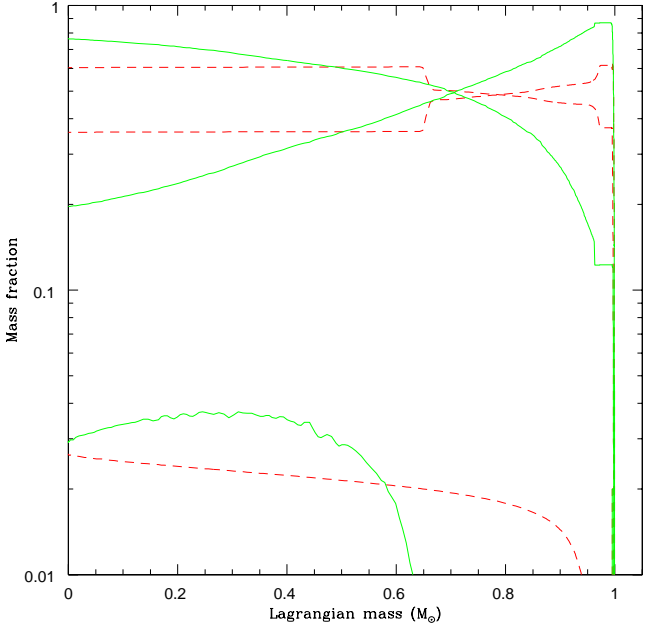}
\caption{Same  as  Fig.~\ref{fig1}  but  for a  diffusion  coefficient
  artificially increased by factor of five.}
\label{fig2}
\end{figure}

Noticeable changes  in the chemical abundance  distribution take place
as the  white dwarf evolves.  During the evolutionary stages  prior to
core crystallization, abundance changes in the white dwarf core result
from  $^{22}$Ne sedimentation.  As  demonstrated by  \citet{del02} and
\citet{gar08},  $^{22}$Ne sedimentation  is a  very slow  process that
influences the evolution of white  dwarfs only after long enough times
have elapsed.  During the evolutionary  stages where most of the white
dwarf  remains  in  a  liquid  state, this  process  causes  a  strong
depletion  of  $^{22}$Ne in  the  outer region  of  the  core, and  an
enhancement of its abundance in  the central regions of the star. This
can be noted  from Fig. \ref{fig1}.  When most of  the white dwarf has
crystallized ---  at $\approx 6$  Gyr --- there remains no trace of $^{22}$Ne 
in the outer parts of the  core. This is more noticeable in the case
of a more  efficient diffusion, as shown in  Fig. \ref{fig2}. 
If  we multiply  the  (rather uncertain)  diffusion  coefficient by  a
factor of $5$,  we indeed find that $^{22}$Ne diffuses so  deep into the core
that  no trace of  this element  is found  in layers  even as  deep as
0.4~M$_{\sun}$ below the stellar surface.

Finally, we note that  the carbon-oxygen  distribution  becomes strongly
modified  by   the  time  the  star  has   ended  its  crystallization
process. This  is a result  of the carbon-oxygen phase  separation and
the induced mixing episodes in the outer liquid layers that take place
upon crystallization  \citep[see][]{ren10}. In particular,  during the
crystallization  process, the  oxygen abundance  in  the crystallizing
region   increases,   and   the   overlying  liquid   mantle   becomes
carbon-enriched  as a result  of mixing  induced by  a Rayleigh-Taylor
instability at  the region above  the crystallized core. Once the
crystallization process is completed, the chemical profiles of all the
elements become frozen, including that of $^{22}$Ne.

\section{Delayed detonations of Chandrasekhar-mass white dwarfs}
\label{sectddt}

Here we report  the results of the explosion of  a massive white dwarf
using the chemical  profiles obtained at the beginning  and the end
of the cooling process. The  pre-supernova model is a cool, isothermal,
Chandrasekhar-mass white dwarf built in hydrostatic equilibrium with a
central density $\rho_\mathrm{c}=3\times10^9$~g~cm$^{-3}$. The central
1~M$_{\sun}$ has  the same  chemical composition as  the carbon-oxygen
core obtained in  the cooling sequences, while the  composition of the
envelope of  mass $M_\mathrm{Ch}-1$~M$_{\sun}$ is the same  as that of
the outermost shell of the carbon-oxygen  core.  As we wish to maximize
the impact  of the chemical profiles  on the supernova  outcome, we do
not take  into account here convective mixing during  the pre-supernova
carbon simmering phase.

The present models are  based on the delayed-detonation (DDT) paradigm
\citep{kho91},  in which  thermonuclear combustion  initially proceeds
through  a subsonic  deflagration until  it  makes a  transition to  a
supersonic     detonation     wave.     The    location     of     the
deflagration-detonation  transition  is  usually parametrized  by  its
density,   $\rho_\mathrm{DDT}$.   The   supernova  hydrodynamics   and
nucleosynthetic (post-processing)  codes we use are  the same as
in \citet{bra96} and \citet{bad03}.
We sketch here, for completeness, the method used by the hydrodynamics code to simulate DDT models
and calculate the nuclear energy generation rate. 
In the DDT models, the flame propagates initially as a deflagration, with a velocity fixed at 
a constant fraction, 3\%, of the local sound velocity. When the flame density {\sl ahead} of the
flame reaches the prescribed transition density, $\rho_\mathrm{DDT}$, 
the flame front is accelerated artificially to a large fraction
of the sound speed, resulting in the subsequent formation of a detonation. Changes in the chemical
composition are followed during the propagation of the
burning front in the deflagration and detonation modes using 
an $\alpha$-network from $^4$He to $^{28}$Si plus the
conversion of $^{28}$Si to $^{56}$Ni in a single step.
When the temperature of a mass shell exceeds $5.5\times10^9$~K, nuclear statistical
equilibrium (NSE) is assumed. 
Once attained, NSE is maintained as long as the temperature remains above $2\times10^9$~K,
providing a nuclear energy generation rate accurate enough for the hydrodynamical simulations. 
Weak interactions during NSE determine the evolution of the electron mole number, $Y_\mathrm{e}$,
\begin{equation}
\frac{\mathrm{d} Y_\mathrm{e}}{\mathrm{d} t} = \Sigma_i \lambda_i Y_i\,,
\end{equation} 
\noindent where $\lambda_i$, accounting for all kind of weak interactions, and the 
molar fractions, $Y_i$, are set by the NSE equations.
The final nucleosynthesis was computed separately with the nuclear reaction
network described by \citet{bra93}, with updated reaction rates taken from the REACLIB
compilation \citep{rau00}, using the temperature and
densities of each mass zone provided by the hydrocode.

Bolometric light curves were obtained by means
of  the  code  described   in  \citet{bra93,bra96}.  In  general,  the
bolometric light curves obtained with this code during the pre-maximum
and maximum phases are in fairly good agreement with those computed by
directly     solving      the     radiative     transfer     equations
\citep[see][]{hoe93}.

Table~\ref{tab1}  gives details  of  the models  we  have computed  so
far. Each model  is characterized by two parameters:  its cooling time
(age),    and   the   deflagration-detonation    transition   density,
$\rho_\mathrm{DDT}$.   This density is  the main  unknown in the DDT SNIa
models.   In  one-dimensional  calculations such as  those  reported  here,
$\rho_\mathrm{DDT}$    is   usually    a   free    parameter.    While
multi-dimensional SNIa  models may be able to emulate the physical dependences
of  $\rho_\mathrm{DDT}$,  in  principle, state-of-the-art  models  are
still nowadays not  able to do it \citep[for a  recent attempt in this
direction  see, e.g.][]{jac10}. Thus,  we repeated the
simulations  for several  values of  $\rho_\mathrm{DDT}$, as  shown in
Table~\ref{tab1}.  In  some calculations,  we used  initial models
resulting from using a diffusion coefficient of $^{22}$Ne during white
dwarf  cooling  artificially increased  by  a  factor of five. For  each
combination of  $\rho_\mathrm{DDT}$ and diffusion  coefficient, we 
computed two models, one belonging to the beginning of crystallization
of the white dwarf (``young''  model) and another at an advanced state
of crystallization ($\sim95-96$\%, ``old'' model).

\begin{table}[t]
\caption{\label{tab1} Results of Chandrasekhar-mass explosions.}
\centering
\begin{tabular}{llcccc}
\hline\hline
Age\tablefootmark{a} & $\rho_\mathrm{DDT}$ &
$M(^{56}\mathrm{Ni})$ & $M(\mathrm{IME})$\tablefootmark{b} & $K$\tablefootmark{c} &
$m_\mathrm{max}$\tablefootmark{d} \\
(Gyr) & (g$~$cm$^{-3}$) & (M$_\odot$) & (M$_\odot$) & ($10^{51}$~erg) & (mag) \\
\hline
0.6 & $1.6\times10^7$ & 0.40 & 0.67 & 1.26 & -18.78 \\
8.0 & $1.6\times10^7$ & 0.43 & 0.70 & 1.37 & -18.84 \\
1.0\tablefootmark{e} & $1.6\times10^7$ & 0.41 & 0.67 & 1.27 & $-18.79$ \\
9.6\tablefootmark{e} & $1.6\times10^7$ & 0.43 & 0.69 & 1.38 & $-18.88$ \\
1.0\tablefootmark{e} & $3.0\times10^7$ & 0.71 & 0.42 & 1.42 & $-19.45$ \\
9.6\tablefootmark{e} & $3.0\times10^7$ & 0.74 & 0.42 & 1.51 & $-19.50$ \\
0.6 & $2.4\times10^7$\tablefootmark{f} & 0.59 & 0.52 & 1.37 & $-19.23$ \\
8.0 & $3.7\times10^7$\tablefootmark{g} & 0.83 & 0.34 & 1.54 & $-19.63$ \\
\hline
\end{tabular}
\tablefoot{All    models:   central   density,    $\rho_\mathrm{c}   =
  3.0\times10^9$~g$~$cm$^{-3}$.\\
\tablefoottext{a}{Time since the beginning of the cooling sequence.\\}
\tablefoottext{b}{Synthesized mass of intermediate-mass elements, from
  Si to Ca.\\}
\tablefoottext{c}{Kinetic energy of the ejecta.\\}
\tablefoottext{d}{Bolometric magnitude at maximum.\\}
\tablefoottext{e}{Model  with the  diffusion coefficient  of $^{22}$Ne
  increased by a factor $5$.\\}
\tablefoottext{f}{Deflagration-to-detonation     transition    density
  estimated  from   the  chemical  composition  at   the  flame  front
  (Eq. \ref{eq1}): $X(^{12}\mathrm{C}) = 0.36$, $X(^{22}\mathrm{Ne}) =
  0.020$.\\}
\tablefoottext{g}{Deflagration-to-detonation     transition    density
  estimated  from the chemical  composition at  the flame  front using
  Eq. (\ref{eq1}): $X(^{12}\mathrm{C}) = 0.24$, $X(^{22}\mathrm{Ne}) =
  0.022$.}  
}
\end{table}

\subsection{Detonating at a fixed $\rho_\mathrm{DDT}$}

The first  two rows of  Table~\ref{tab1} give the mass  of radioactive
$^{56}$Ni,  the  mass   of  intermediate-mass  elements  (IME,  mostly
composed  of Si  and S),  the kinetic  energy of  the ejecta,  and the bolometric
magnitude of the supernova at  maximum, for two models computed at the
beginning  and the  end  of  the white  dwarf  cooling sequence,  with
$\rho_\mathrm{DDT}=1.6\times10^7$~g$~$cm$^{-3}$. The results are quite
insensitive  to  the  age  of  the  pre-supernova,  the  difference  in
brightnesses being  $\sim0.06$~mag.  The  differences in both the  yield of
IMEs and the kinetic energy are correspondingly small, on the order
of 5\% and 9\%, respectively.

\subsubsection{Enhanced diffusion of $^{22}$Ne}

The  third and  fourth rows  of Table~\ref{tab1}  give the  results of
models  identical to  the ones  just  discussed but  for the  enhanced
diffusion  coefficient of  $^{22}$Ne during  white dwarf  cooling. The
results are  nearly identical  to those  obtained with  the nominal
diffusion  coefficient.   Thus,  any  uncertainty   in  the  diffusion
coefficient  does not  translate into  appreciable differences  in the
observational properties  of SNIa.  These  conclusions are independent
of   the  value   of   $\rho_\mathrm{DDT}$  used   in  the   supernova
calculations, as  can be seen in the  next two rows where  we show the
results for $\rho_\mathrm{DDT}=3.0\times10^7$~g$~$cm$^{-3}$.

\begin{figure}[t]
\centering
   \includegraphics[width=9 cm]{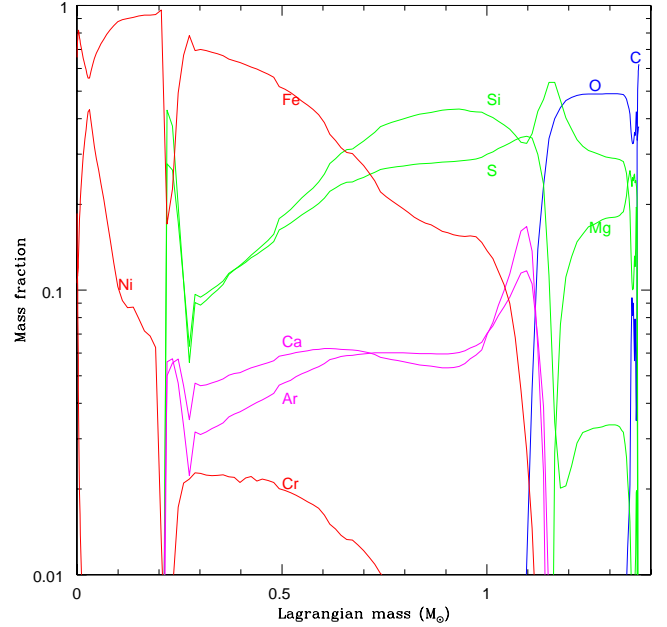}
\caption{Final  chemical  composition  of  the ejecta  for  the  model
  detonated  at the  beginning of  the  cooling process  with a  fixed
  $\rho_\mathrm{DDT}   =  1.6\times10^7$~g$~$cm$^{-3}$   and  enhanced
  diffusion.  The  abundances shown reflect  the elemental composition
  after  radioactive disintegrations.  See the  online edition  of the
  journal for a color version of the figure.}
\label{fig3}
\end{figure}

\begin{figure}[tb]
\centering
   \includegraphics[width=9 cm]{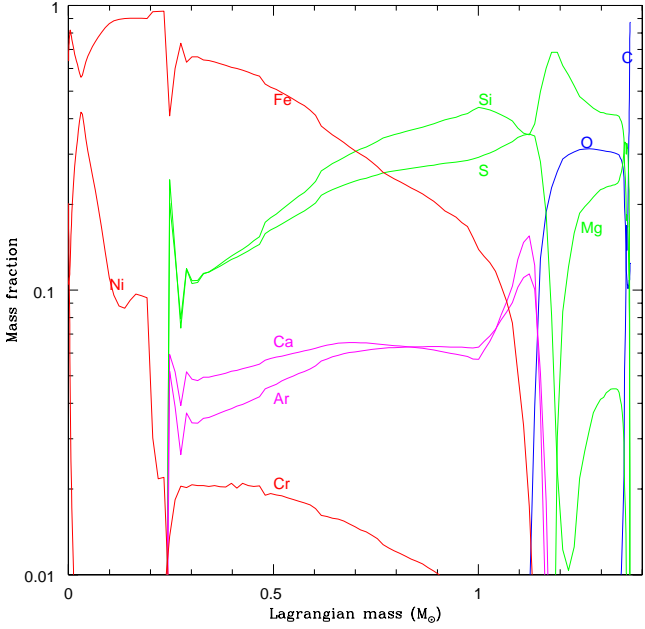}
\caption{Same   as   Fig.~\ref{fig3},   but   at  the   end   of   the
  crystallization phase.}
\label{fig4}
\end{figure}

Figures~\ref{fig3} and \ref{fig4}  show the final chemical composition
of  the ejecta  as a  function  of the  ejecta mass  (as a  Lagrangian
coordinate), for  the two  models with enhanced  diffusion coefficient
and    $\rho_\mathrm{DDT}=1.6\times10^7$~g$~$cm$^{-3}$.     The   main
difference between  the initial  models (see Fig.~\ref{fig2})  lies in
the concentration of $^{22}$Ne  in the central 0.6~M$_{\sun}$, which leads
to a larger $^{22}$Ne mass  fraction ($\sim50$\%) in the oldest model.
However, the chemical profiles of the ejecta of both models are nearly
the same below this Lagrangian mass.  Between $\sim0.6$~M$_{\sun}$ and
1~M$_{\sun}$, the  oldest model (8.0~Gyr age) has  slightly larger mass
fractions of Si and S and less Fe and Cr, although the differences are
quite modest. The reason for these differences is the following. As the
thermonuclear wave propagates through the central regions of the white
dwarf, it completely incinerates the material up to  a composition in
nuclear  statistical equilibrium.   The oldest  model having  a
larger  central neutron  excess  (larger $^{22}$Ne  mass fraction) 
releases less  energy during the incineration of carbon and  oxygen in reaching
NSE, which causes a slightly slower expansion  during the initial
phases of  the explosion, and higher densities at  the position of
the flame.  As a consequence,  the subsequent combustion of the layers
above a $\sim0.6$~M$_{\sun}$  Lagrangian mass is able to  reach a more
advanced combustion stage than if the expansion had been faster
(as in the youngest model, of  0.6~Gyr age), and more Fe and less IMEs
are produced in  these layers. 

\subsection{Transition density function of the chemical composition}

We now speculate whether $\rho_\mathrm{DDT}$  is a
function       of      the       local       chemical      composition
\citep{cha07,woo07b,bra10,jac10}\footnote{The models discussed in this
  section   are  based  on   the  nominal   value  of   the  diffusion
  coefficient.}.  In considering this  scenario,  we scaled  the  transition
density as  a function  of the local  chemical composition  as
\citep[see][for details]{bra10}

\begin{equation}
\rho_\mathrm{DDT}\propto X(^{12}\mathrm{C})^{-1.3}\left(1+129\eta\right)^{-0.6}\,,
\label{eq1}
\end{equation}

\noindent  where  the  neutron  excess,  $\eta=1-2Y_\mathrm{e}$,  is  related  to  the
$^{22}$Ne mass fraction by

\begin{equation}
 \eta=\frac{X(^{22}\mathrm{Ne})}{11}
\end{equation}

\begin{figure}[t]
\centering
   \includegraphics[width=9 cm]{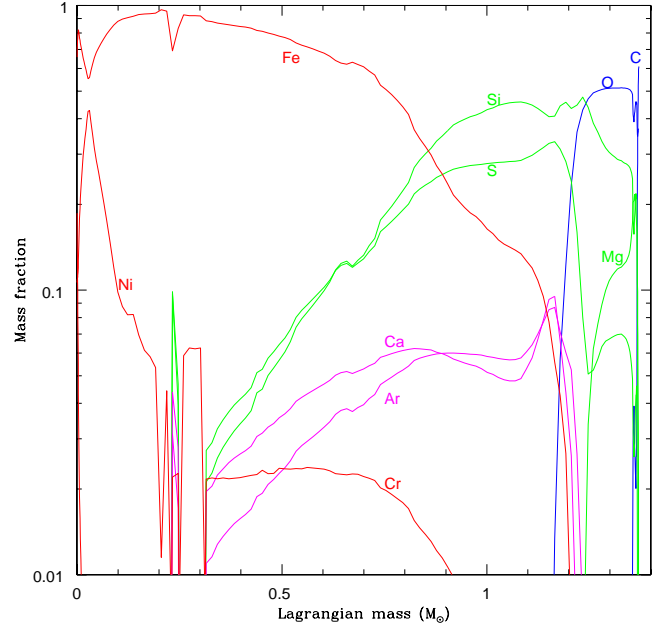}
\caption{Chemical  composition  of  the  ejecta  for  the  model  with
  $\rho_\mathrm{DDT}$ function  of the mass fractions  of $^{12}$C and
  $^{22}$Ne  as given  by  Eq.  (\ref{eq1}) at  the  beginning of  the
  cooling process: $X(^{12}\mathrm{C}) = 0.36$, $X(^{22}\mathrm{Ne}) =
  0.020$ (see Table~\ref{tab1}).}
\label{fig5}
\end{figure}

\begin{figure}[t]
\centering
   \includegraphics[width=9 cm]{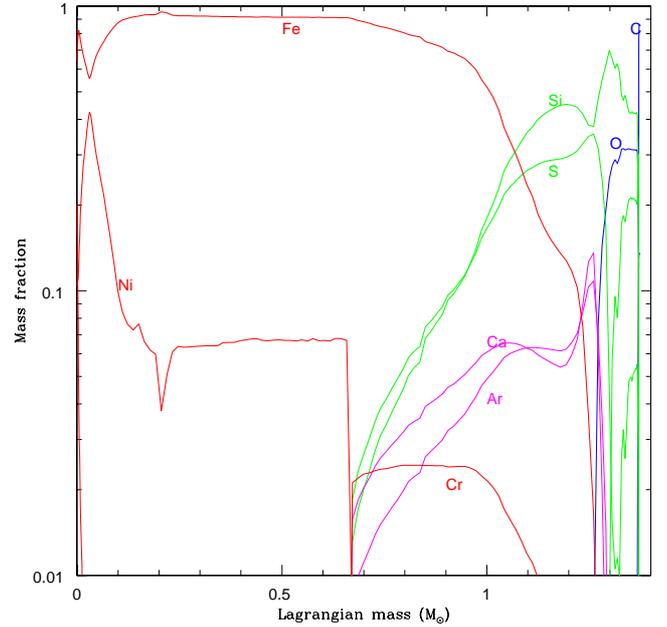}
\caption{Chemical  composition  of  the  ejecta  for  the  model  with
  $\rho_\mathrm{DDT}$ function  of the mass fractions  of $^{12}$C and
  $^{22}$Ne  as  given   by  Eq.   (\ref{eq1})  at  the   end  of  the
  crystallization     phase:      $X(^{12}\mathrm{C})     =     0.24$,
  $X(^{22}\mathrm{Ne}) = 0.022$ (see Table~\ref{tab1}).}
\label{fig6}
\end{figure}

Equation~(\ref{eq1})  implies  that a  reduction  in  the carbon  mass
fraction  translates  into  a higher  $\rho_\mathrm{DDT}$,  while  an
increase in the $^{22}$Ne mass  fraction has the opposite effect.  The
results  we  obtained  are   given  in  the  last  two  rows  of
Table~\ref{tab1},  and in Figs.~\ref{fig5}  and \ref{fig6},  where the
final  chemical composition  of the  ejecta  is shown  for the  models
corresponding to both the beginning and  the end of the white dwarf cooling
process.

Allowing $\rho_\mathrm{DDT}$  to be  a function of  the local  (at the
position  of  the flame  front)  chemical  composition  makes a  great
difference. The transition from deflagration to detonation takes place
at a Lagrangian mass $\sim0.22$~M$_{\sun}$\footnote{The location of the
  DDT  can be  identified  in Figs.~\ref{fig3}  to  \ref{fig6} with the
  sudden  drop in the  Ni profile  slightly above  0.2~M$_{\sun}$. This
  discontinuity  is a  numerical  artifact of  the  procedure used  to
  transmute  a  deflagration  front  into  a  detonation  wave.}.  The
abundances of  $^{12}$C and  $^{22}$Ne at this  location are  0.36 and
0.020, respectively,  for the youngest  model, and 0.24 and  0.022 for
the oldest model (see Fig.~\ref{fig1}). The main effect comes from the
reduction  in the  carbon  mass  fraction with  age, which causes an
increase of $\sim63$\% in $\rho_\mathrm{DDT}$. The modest variation in
the  neon abundance  at  the  flame front  only  makes the  transition
density decrease  by  $\sim1$\%.  Finally,   the  difference  in
$\rho_\mathrm{DDT}$  as a  function of  the progenitor  age translates
into a SNIa brightness difference of 0.40~mag.

In  these calculations,  the production  of IMEs  by SNIa  is strongly
affected by the age of the progenitor, since it is reduced by a factor of
$\sim0.65$ for  the oldest model.  As can be seen  in Figs.~\ref{fig5}
and \ref{fig6},  the larger $\rho_\mathrm{DDT}$ is, the  more extended
the Fe production  region and the more squeezed the region rich in
IMEs is.

\section{Explosion of sub-Chandrasekhar white dwarfs}
\label{sectsub}

The          sub-Chandrasekhar          model         of          SNIa
\citep{woo80,nom82,woo86b,ibe91,lim91,liv91,woo94b,gar99}           was
challenged 14  years ago  because its
spectra  was  excessively  blue compared to SNIa  observations
\citep{hk96}. This disagreement had its origin in the assumption that a
thick   helium  layer   was   necessary to  detonate   a
sub-Chandrasekhar  white  dwarf.   However,  there  has  been  renewed
interest in this  kind of explosion because of the possibility
that   they  produce   a   new  class   of  underluminous   supernovae
\citep{bil07}, and the  finding that a  white dwarf  detonation is
possible  for  lower helium  shell  masses  than previously  thought
\citep{fin10}.  This last point  allows a  better agreement  with SNIa
optical  properties  \citep{kro10}  but  the models  still  show  some
deficiencies, particularly in terms of their lack of a central region  rich in stable
Fe-group elements,  as demanded by infrared observations  of SNIa \citep{hoe04}. One
possible  solution  for this  deficiency,  proposed by  \citet{kro10},
might be a higher neutron excess in  the central regions  of the
white dwarf caused by gravitational settling  of $^{22}$Ne
during white dwarf cooling. We test this hypothesis in the following.

The   initial   models  used   in   our  sub-Chandrasekhar   explosion
calculations  are  composed  of  a  carbon-oxygen core  and  a  helium
envelope.   The  core  has   the  same   chemical  structure   as  the
carbon-oxygen core  in the 1~M$_{\sun}$  white dwarf evolved  with the
{\tt  LPCODE}  code (Sect.~\ref{sectchem}),  taken  at different  ages
since the beginning of the cooling process. The envelope has a mass of
0.080~M$_{\sun}$ and is composed of 100\% He. The mechanical structure
of the white  dwarf that resulted from the  cooling sequences was
readjusted to  reestablish the hydrostatic equilibrium taking
into account  the envelope. As  a result, the central  density changed
from  $\sim3.5\times10^7$~g$~$cm$^{-3}$,  as   given  by  the  cooling
sequences,   to  $\sim5.8\times10^7$~g$~$cm$^{-3}$   in   our  initial
models.   The  density   at  the   base   of  the   He  envelope   was
$\sim2.1\times10^6$~g$~$cm$^{-3}$.

The models  we computed  are summarized in  Table~\ref{tab2}. The
models were calculated at  two different ages of  the progenitor
white dwarf,  and the  initial He detonation was started  at two
different locations: just  at the base of the  helium envelope, and at
an altitude  of 70~km above this.  We also computed a reference
model in which  the composition of the core is  uniform and has equal
mass fractions of carbon and oxygen (first row in Table~\ref{tab2}).

The precise point  at which He detonates is  not known with precision
because of unresolved convective  motions during the  first stages  of He
burning at the base of the  envelope, so the above procedure allows us to
evaluate the impact of the chemical differentiation of the white dwarf
in two  extreme conditions.  Detonating helium at  different altitudes
can have consequences for the outcome  of the inward shock wave that is
launched  into the  white  dwarf core  \citep{liv91}.  If He  ignition
occurs at  a high  enough altitude above  the core, the  He detonation
wave has time to develop well before it arrives at the core and it can
be  strong enough  then to directly induce an  inwardly moving  carbon
detonation.  This  possibility depends  as  well  on  the carbon  mass
fraction at the outermost part  of the core, which in turn depends  on the age of
the progenitor white dwarf.

\begin{table}[t]
\caption{\label{tab2} Results of sub-Chandrasekhar explosions.}
\centering
\begin{tabular}{cccccc}
\hline\hline
Age\tablefootmark{a} & $h_\mathrm{ig}$\tablefootmark{b} & $M(^{56}\mathrm{Ni})$ &
$M(\mathrm{Fe,sta})$\tablefootmark{c} & $M(\mathrm{IME})$ & $K$\\
(Gyr) & (km) & (M$_\odot$) & (M$_\odot$) & (M$_\odot$) & ($10^{51}$~erg) \\
\hline
-\tablefootmark{d} & 0 & 0.70 & 0.055 & 0.22 & 1.39 \\
0.6 & 0 & 0.69 & 0.054 & 0.23 & 1.34 \\
8.0 & 0 & 0.68 & 0.058 & 0.24 & 1.35 \\
0.6 & 70 & 0.63 & 0.054 & 0.28 & 1.31 \\
8.0 & 70 & 0.64 & 0.057 & 0.30 & 1.34 \\
\hline
\end{tabular}
\tablefoot{All    models:   Mass    of    the   carbon-oxygen    core,
  $M_\mathrm{core}   =  1$~M$_{\sun}$;   mass  of   the   He  envelope
  $M_\mathrm{env}     =     0.080$~M$_{\sun}$;    central     density,
  $\rho_\mathrm{c} = 5.8\times10^7$~g$~$cm$^{-3}$; density at the base
  of       the      He      envelope,       $\rho_\mathrm{base}      =
  2.1\times10^6$~g$~$cm$^{-3}$.\\
\tablefoottext{a}{Time since the beginning of the cooling sequence.\\}
\tablefoottext{b}{Altitude at which the He detonation starts.\\}
\tablefoottext{c}{Ejected mass of {\sl stable} Fe-group nuclei.\\}
\tablefoottext{d}{Reference model,  with a homogeneous  composition of
  the core of 49\% carbon and oxygen plus 2\% $^{22}$Ne.\\}
}
\end{table}

\subsection{Edge-lit detonation of carbon}

We  explored  the  conditions  for the  formation  of  a  stable
self-sustained  carbon detonation  induced by  a He  detonation,  as a
function of the  carbon mass fraction and the altitude  at which He is
ignited. A  similar analysis was performed by  \citet{gar99}, but they
explored a range of $^{12}$C  mass fractions below 0.5, while those we
are interested  in are  much larger ($X(^{12}\mathrm{C})=0.61$  in our
young  progenitor  model,  and  $X(^{12}\mathrm{C})=0.86$ in  the  old
progenitor  model).   \citet{gar99}  found  that  the  probability  of
a self-sustained detonation occurring in the core is very sensitive
to the carbon mass fraction.

In  the  present  numerical  experiments, we applied  the  same
methodology  adopted in  \citet[][and references  therein]{bra09a}. We
followed   the  hydrodynamical  and  nuclear   evolution  of  an
isothermal uniform density sphere consisting of a central ball made of
99\% $^{4}$He and  1\% $^{12}$C, surrounded by a  much higher mass of
the same  chemical composition as the outermost  layer of the
carbon-oxygen core of  the white dwarf, produced by the cooling
sequences.  The radius  of  the central  ball  plays the  role of  the
altitude at which  He ignites in the sub-Chandrasekhar  models. The He
detonation  was started  by incinerating  a central  region containing
between 1\% and 8\% of the mass of the central helium-rich ball.

\begin{table}[b]
\caption{\label{tab3} Induced detonation of carbon.}
\centering
\begin{tabular}{ccrc}
\hline\hline
$\rho$ & $X(^{12}\mathrm{C})$\tablefootmark{a} & Altitude\tablefootmark{b} &
Detonation? \\ 
(g$~$cm$^{-3}$) & & (km) & \\
\hline
$6\times10^6$ & 0.50 & 5 & no \\
$6\times10^6$ & 0.61 & 5 & no \\
$6\times10^6$ & 0.86 & 5 & yes \\
$6\times10^6$ & 0.50 & 10 & {\sl weak} \\
$6\times10^6$ & 0.61 & 10 & yes \\
$6\times10^6$ & 0.86 & 10 & yes \\
$6\times10^6$ & 0.50 & 20 & yes \\
$2\times10^6$ & 0.50 & 50 & no \\
$2\times10^6$ & 0.61 & 50 & no \\
$2\times10^6$ & 0.86 & 50 & no \\
$2\times10^6$ & 0.50 & 100 & no \\
$2\times10^6$ & 0.61 & 100 & {\sl weak} \\
$2\times10^6$ & 0.86 & 100 & {\sl weak} \\
$2\times10^6$ & 0.50 & 200 & {\sl weak} \\
$2\times10^6$ & 0.61 & 200 & {\sl weak} \\
$2\times10^6$ & 0.86 & 200 & {\sl weak} \\
\hline
\end{tabular}
\tablefoot{
\tablefoottext{a}{Carbon  mass  fraction  surrounding the  central  He
  ball. In  all calculations, the  oxygen mass fraction in  this region
  was set to $X(^{16}\mathrm{O}) = 1 - X(^{12}\mathrm{C})$.\\}
\tablefoottext{b}{Radius of the central ball made of 99\% $^{4}$He and
  1\% $^{12}$C.}  
}
\end{table}

The  results   of  these  numerical  experiments   are  summarized  in
Table~\ref{tab3}, where  we have investigated the  induction of carbon
detonation at two densities: $2\times10^6$~g$~$cm$^{-3}$, which matches
the density at the base of the helium envelope in our 1.080~M$_{\sun}$
sub-Chandrasekhar models,  and $6\times10^6$~g$~$cm$^{-3}$, which would
be representative of models with  a higher total mass.  Here, a ``weak
detonation'' means a carbon detonation in which the temperature is not
high  enough  to  burn  oxygen directly  \citep[see][]{gar99}.   In  a
sub-Chandrasekhar  explosion, such an  edge-lit  weak detonation  will
strengthen  as it  moves  into denser  regions, reaching  increasingly
higher temperatures and allowing the completion of burning  sequences from
explosive oxygen burning until NSE. The successful
launching of a carbon detonation  wave is found to depend strongly on the density
and  the  altitude of  He  ignition, and  weakly  on  the carbon  mass
fraction. For instance,  for a density of $2\times10^6$~g$~$cm$^{-3}$,
the  minimum altitude  at which  a carbon  detonation develops  in our
simulations  is $200$~km  for a  composition of  50\%:50\%  carbon and
oxygen, and  only $100$~km for  the two compositions richer  in carbon
that  we explored.   For  the highest  density  we used,
$6\times10^6$~g$~$cm$^{-3}$,  and  an   altitude  of  5~km,  a  carbon
detonation  was obtained only  for the  largest carbon  mass fraction,
$X(^{12}\mathrm{C})=0.81$. Thus, the age  of the progenitor system can
determine the  kind  of  dynamical  event
following  He  ignition in a sub-Chandrasekhar  white  dwarf, i.e. either  a
central carbon detonation or a shell carbon detonation.

\subsection{Explosion models}

The effects of white dwarf crystallization and $^{22}$Ne sedimentation on  
sub-Chandrasekhar  explosions   can  be   seen  in
Table~\ref{tab2}. First, comparing the three models that ignited He at
the base of the envelope (first three rows, with $h_\mathrm{ig}=0$) we
see that  the impact  of chemical differentiation  is minimum,  on the
order of a few percent in $^{56}$Ni and IMEs yields and in the kinetic
energy.  Second, comparing  the two models with $h_\mathrm{ig}=70$~km,
we observe the same lack of dependence of the outcome of the explosion
on the age of the progenitor system.

\begin{figure}
\centering
   \includegraphics[width=9 cm]{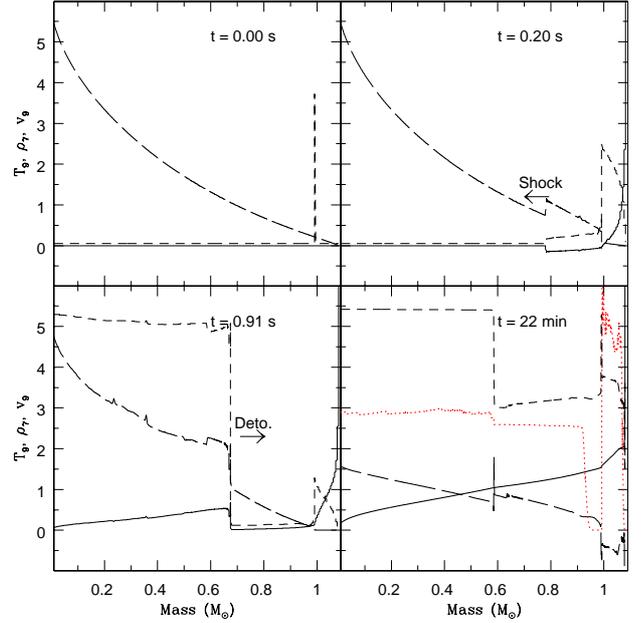}
\caption{Development  of the  detonation of  a  sub-Chandrasekhar mass
  white  dwarf   that  explodes  at  the  beginning   of  the  cooling
  process.  From left  to right  and top  to bottom,  the  first three
  panels  show  the  profiles  of  temperature in  units  of  $10^9$~K
  (short-dashed  line),  density   in  units  of  $10^7$~g$~$cm$^{-3}$
  (long-dashed  line), and velocity  in units  of 10,000~km$~$s$^{-1}$
  (solid   line),  at   three  times   since  the   beginning   of  He
  detonation.  In this  model,  the He  detonated  at the  base of  the
  envelope,  so it  was unable  to produce  an edge-lit  detonation of
  carbon (see Tables~\ref{tab2} and \ref{tab3}). Instead, an initially
  weak inwardly  moving shock traveled  through the white  dwarf (second
  panel). Shortly  before reaching the center,  the shock strengthened
  enough  to  produce  a  carbon detonation that  traveled  outwards
  processing most  of the matter on top (third  panel). The  last panel
  shows   the  final  profiles   of  $\log(T)$   (short-dashed  line),
  $\log(10^5\rho)$  (long-dashed  line),  velocity (solid  line),  and
  $50X(\mathrm{Fe,stable})$,  where  $X(\mathrm{Fe,stable})$ is the  mass
  fraction  of   stable  Fe-group   nuclei  (red  dotted   line).  The
  discontinuity in the thermal profile in the last panel is due to the
  local  deposition   of  the  photons  emitted   in  the  radioactive
  disintegration of $^{56}$Ni.}
\label{fig7}
\end{figure}

\begin{figure}
\centering
   \includegraphics[width=9 cm]{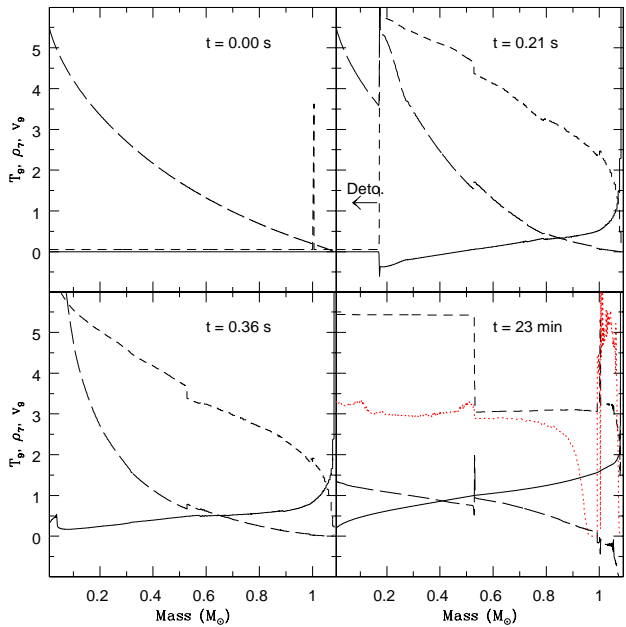}
\caption{Development  of the  detonation of  a  sub-Chandrasekhar mass
  white  dwarf  that  explodes  at  the  end  of  the  crystallization
  phase.  The  meaning of  the  different curves  is  the  same as  in
  Fig.~\ref{fig7}. In  this model, the  He detonated at an  altitude of
  70~km from  the base of the  envelope, and launched  a second inwardly
  moving carbon detonation  (see Tables~\ref{tab2} and \ref{tab3}), as
  can  be seen  already in  the  second panel.  The carbon  detonation
  processed all the matter down to  the center of the white dwarf and,
  thereafter, sent  a weak shock  wave outwards (visible in  the third
  panel at a Lagrangian mass of $\sim0.5$~M$_{\sun}$).}
\label{fig8}
\end{figure}

A further comparison can be established between a young model in which
the edge-lit detonation of  carbon fails (and, later, carbon detonates
at  the center of  the white  dwarf), and  an old  model in  which the
edge-lit  detonation  of  carbon  is  successful.   Figures~\ref{fig7}
(young  model, second  row  in Table~\ref{tab2})  and \ref{fig8}  (old
model, last row  in Table~\ref{tab2}) show the evolution  of the white
dwarf in both cases. Even though the history of the explosion is quite
sensitive to  the location of  carbon detonation, the final  result is
not so different. In the last panel of Figs.~\ref{fig7} and \ref{fig8},
we show  the final mechanical,  thermal, and velocity profiles  of the
ejecta,  together with  the profile  of  the mass  fraction of  stable
Fe-group elements  (red dotted  line). The  concentration of
stable Fe-group elements  is clearly insensitive to the age  of the progenitor
system of the exploding sub-Chandrasekhar white dwarf.

\section{Conclusions}
\label{sectcon}

We  have   investigated  the   sensitivity  of  SNIa   energetics  and
nucleosynthesis to  realistic chemical  profiles based on  white dwarf
cooling sequences calculated by  \citet{alt10}.  The cooling
sequences provide  a link between  the white dwarf  chemical structure
and the age of  the supernova progenitor system. Neither gravitational
settling  of $^{22}$Ne  nor chemical  differentiation of  $^{12}$C and
$^{16}$O during  white dwarf crystallization have a  sizable impact on
the properties  of SNIa,  unless there is  a direct dependence  of the
flame properties on chemical  abundances. If the density of transition
from  deflagration  to detonation  in  SNIa  did  not depend  on  the
chemical composition,  the variation in the supernova  magnitude with
age  produced  by  chemical  differentiation  would  be  as  small  as
$\sim0.06$ magnitudes for an age difference of $\sim7.4$~Gyr. 
We emphasize that these results have been obtained by neglecting mixing during the
pre-supernova carbon simmering phase. If this mixing process is efficient, it will erase
any trace of the chemical separation achieved during white dwarf cooling, thus the
chemical separation will leave no imprint on the supernova properties. Our results therefore
represent the maximum possible effect that can be expected from the gravitational settling of
$^{22}$Ne and $^{16}$O.

If the density  of transition from deflagration to  detonation in SNIa
is a function of the chemical  composition, and if we neglect convective
mixing during the pre-supernova  carbon simmering phase, the difference
in maximum magnitude between a  white dwarf exploding at the beginning
of  the   cooling  process  and  another  one   that  has  experienced
substantial  crystallization can  be as  large as  $\sim0.4$~mag.  For
this variation in magnitude to take place, it suffices that the central
part of the white dwarf has crystallized (22\% in mass in our models),
so that the transition from  deflagration to detonation takes place in
a region that  has been depleted in carbon.   For a 1~M$_{\sun}$ white
dwarf, the  crystallization of the central  22\% in mass  occurs at an
age of only 1.25~Gyr.

The  physics  of  $^{22}$Ne  sedimentation  is  nowadays  sufficiently
well-known  that  the  calculations  presented in  this  paper  cannot
experience much  variation.  We  have also explored the  impact of
using   an   unrealistically   large  diffusion   coefficient,   which
nevertheless had a  very small impact on the  observable properties of
SNIa explosions.   We stress that  to obtain a large  concentration of
$^{22}$Ne in the  central layers of a white dwarf  the star would have
to be made of pure carbon, which is not applicable in the more general
case of  carbon-oxygen white dwarfs. Thus, our  results concerning the
lack  of imprint  of the  $^{22}$Ne  sedimentation on  SNIa are  quite
robust.

In the light  of the present results, the only  possible ways in which
chemical differentiation might affect the thermonuclear explosion of a
white  dwarf are  either  through a composition dependent 
density of transition from a  deflagration to a detonation, or through
composition dependent thermonuclear runaway conditions (usually taken to be
the initial conditions  for SNIa
simulations).  Linking explosion models  to the  last stages  of white
dwarf evolution remains one of the main challenges of SNIa theory.

Finally, we highlight the importance  of developing realistic pre-supernova
evolutionary  calculations to  understand the  constraints  imposed by
observations  on SNIa  models. In the case  of sub-Chandrasekhar explosions, we have  proven that
the
sedimentation of  $^{22}$Ne does not efficiently  increase the neutron
excess  in the  central  region of  the  white dwarf  prior to  the
explosive event. Thus, this sedimentation cannot be responsible for the
production  of the stable  Fe-rich core inferred from observations of
SNIa.  However, the  crystallization  of the  white  dwarf entails  an
increase in the  carbon  abundance  in the  outermost  layers of  the
carbon-oxygen white  dwarf core, which may have consequences  for the
formation of a double  detonation (outwards in He-rich matter, inwards
in the  C-O core) following helium  ignition close to the  base of the
helium mantle.

\begin{acknowledgements}
This  work  was partially  supported  by  the  AGAUR, by  MCINN  grant
AYA2008--04211--C02,  by  the   European  Union  FEDER  funds,  by
AGENCIA: Programa de Modernizaci\'on Tecnol\'ogica BID 1728/OC-AR, and
by PIP 2008-00940  from CONICET. LGA also acknowledges  a PIV grant of
the AGAUR of the Generalitat de Catalunya.
\end{acknowledgements}

\bibliographystyle{aa}

\end{document}